\documentstyle[12pt,epsf]{article}
\renewcommand{\bar}[1]{\overline{#1}}

\begin{document}

\begin{flushright}
USM-TH-128
\\ CPT-2002/P.4414
\end{flushright}
\bigskip\bigskip

\centerline{
\bf Phenomenological Relation between
Distribution and Fragmentation Functions}

\vspace{22pt} \centerline{\bf Bo-Qiang Ma\footnote{e-mail:
mabq@phy.pku.edu.cn}$^{a}$, Ivan Schmidt\footnote{e-mail:
ischmidt@fis.utfsm.cl}$^{b}$, Jacques Soffer\footnote{e-mail:
Jacques.Soffer@cpt.univ-mrs.fr}$^{c}$, Jian-Jun
Yang\footnote{e-mail: jjyang@fis.utfsm.cl}$^{b,d}$}

\vspace{2pt}

{\centerline {$^{a}$Department of Physics, Peking University,
Beijing 100871, China,}}



{\centerline {$^{b}$Departamento de F\'\i sica, Universidad
T\'ecnica Federico Santa Mar\'\i a,}}

{\centerline {Casilla 110-V, 
Valpara\'\i so, Chile}

{\centerline {$^{c}$Centre de Physique Th$\acute{\rm{e}}$orique,
CNRS, Luminy Case 907,}}

{\centerline { F-13288 Marseille Cedex 9, France}}

{\centerline {$^{d}$Department of Physics, Nanjing Normal
University, Nanjing 210097, China}}

\vspace{4pt}
\begin{center} {\large \bf Abstract}

\end{center}

We study the relation between the quark distribution function
$q(x)$ and the fragmentation function $D_q(z)$ based on a general
form $D_q(x)=C(z)z^{\alpha}q(z)$ for valence and sea quarks. By
adopting two known parametrizations of quark distributions for the
proton, we find three simple options for the fragmentation
functions that can provide a good description of the available
experimental data on proton production in $e^+e^-$ inelastic
annihilation. These three options support the revised
Gribov-Lipatov relation $D_q(z)=z q(z)$ at $z \to 1$, as an
approximate relation for the connection between distribution and
fragmentation functions. The three options differ in the sea
contributions and lead to distinct predictions for antiproton
production in the reaction $ p+p \to \bar{p}+X$, thus they are
distinguishable in future experiments at RHIC-BNL.

\centerline{PACS numbers: 13.87.Fh, 13.60.Hb, 13.65.+i, 13.85.Ni}

\newpage

The quark distribution function $q(x)$ and the quark fragmentation
function $D_q(z)$ are two basic quantities on the structure of
hadrons. It would be very useful if there exist simple connections
between them, so that one can predict the poorly known $D_q(z)$
from the rather well known $q(x)$. Conversely, one could predict
the quark distribution functions of a hadron that cannot be used
as a target from the quark fragmentation to the same hadron. In
fact, some efforts have been made to connect fragmentation
functions with the corresponding quark distributions
\cite{GL,BDM00,BRV00}. There is a so called Gribov-Lipatov
``reciprocity" relation \cite{GL} which connects the distribution
and fragmentation functions in a form
\begin{equation}
D_q(z) \propto q(x). \label{GLR}
\end{equation}
Such relation has been used as a useful ``Ansatz" to model the quark
fragmentation functions based on predictions of quark
distributions functions \cite{GL,MSSY}. Recently, based on
theoretical arguments with some assumptions
\cite{BDM00}, a revised form of the Gribov-Lipatov ``reciprocity" relation, i.e.,
\begin{equation}
D_q(z)=z q(z),
\label{RGLR}
\end{equation}
has been suggested as an approximate relation at $z \to 1$. Thus it is
necessary to check the validity of Eq.~(\ref{GLR}) and/or
Eq.~(\ref{RGLR}) by means of careful phenomenological studies.

The nucleon is a satisfactory laboratory to check the relation
between fragmentation and distribution functions, since we have
data both on the quark distributions of the nucleon from deep
inelastic scattering (DIS) \cite{DISprocess,DYprocess} and on the
fragmentation functions of quark to proton from $e^+e^-$ inelastic
annihilation (IA). Various parametrizations of parton
distributions have been obtained from the DIS experimental data on
the nucleon, with rather high precision
\cite{CTEQ,BSB01,GRV95,MRST00}. We can start with any set of
parton distributions to parametrize the fragmentation functions.
In this paper, we will adopt the CTEQ parametrization (CTEQ6 set
1) \cite{CTEQ} of the parton distributions and for comparison, we
will also use another recent parametrization obtained in the
statistical physical picture, by Bourrely, Soffer and Buccella
(BSB)~\cite{BSB01}. These parametrizations provide reliable
information on the valence and sea quarks and their respective
roles can be studied separately. On the other hand, some
experimental data on proton production in $e^+e^-$ IA are
available~\cite{ALEPH98} and can be used to constrain the shape of
the fragmentation functions of quark to proton. By parametrizing
the fragmentation functions of quark to proton based on a reliable
set of parton distributions and confronting them with the
available experimental data on proton production, we thus have an
effective way to learn about the relation between fragmentation
and distribution functions. For this analysis, we adopt a general
form to relate fragmentation and distribution functions and we
make a distinction between valence and sea quarks, as follows
\begin{equation}
\begin{array}{ll}
D_v(z)=C_v(z) z^{\alpha}q_v(z),\\
D_s(z)=C_s(z) z^{\alpha}q_s(z).
\end{array}
\label{dfR}
\end{equation}
The above forms are always correct, since $C_v(z)$ and $C_s(z)$
are in principle arbitrary. Strictly speaking, there is no way to
discriminate between ``valence" and ``sea" fragmentation, because
both ``valence" and ``sea" quarks $q$ (not $\bar{q}$) can fragment
at a same $z$. Since it is not possible to distinguish whether the
fragmenting quark is ``valence" or ``sea", we should consider
Eq.~(\ref{dfR}) as a phenomenological parametrization for the
fragmentation functions of quarks and antiquarks, as follows
\begin{equation}
\begin{array}{ll}
D_q(z)=D_v(z)+D_s(z),\\
D_{\bar{q}}(z)=D_s(z).
\end{array}
\label{Dqqbar}
\end{equation}

It will be shown in this Letter that, in order to fit the
experimental data of proton production in IA, with fragmentation
functions based on the above parametrizations, we can have very
simple forms of $C_v(z)$ and $C_s(z)$ together with simple values
for $\alpha$. Three options are found to be very good: (1) $C_v=1$
and $C_s=0$ for $\alpha=0$, (2) $C_v=C_s=1$ for $\alpha=0.5$, and
(3) $C_v=1$ and $C_s=3$ for $\alpha=1$. Since the sea quark
contributions are negligible at large $x$, all of the three
options support the revised Gribov-Lipatov ``reciprocity" relation
$D_q(z)=z q(z)$ at $z \to 1$, as an approximate relation for the
connection between distribution and fragmentation functions.
Option 1 corresponds to the case of a suppressed sea, whereas
Option 3 corresponds to the case of an enhanced sea, so they
provide two extreme situations. Thus these three options lead to
different fragmentation pictures related to the respective roles
of valence and sea quarks. We will show that they provide
different predictions for proton/antiproton productions in the
reactions $ p+p \to p (\bar p) +X$, thus they are distinguishable
in future experiments at RHIC-BNL.

The revised Gribov-Lipatov relation Eq.~(\ref{RGLR}) is only known
to be valid near $z \to 1$ and on a certain energy scale $Q^2_0$,
in the leading approximation. Since distribution and fragmentation
functions have different evolution behaviors, it is appropriate to
apply the relation at a starting energy scale  $Q^2_0$, and then
to evolve the fragmentation functions to the experimental scale.
Our goal is to get a good description for the cross section of
proton production in $e^+e^-$ IA \cite{ALEPH98}. In our analysis,
we adopt the initial energy scale $Q^2_0$ of the quark
distributions, i.e., it is $Q_0=1.3$~{GeV} for the CTEQ
parametrization \cite{CTEQ} and $Q_0=2$~{GeV} for the BSB
parametrization \cite{BSB01}. In addition, in the evolution the
gluon fragmentation functions are also needed, and we assume that
they have the same relation with the corresponding gluon
distribution functions as the sea quarks. For the evolution we
adopt the evolution package of Ref.~\cite{BSB01}, modified for the
fragmentation functions.

In the quark-parton model, the differential cross section for the
semi-inclusive proton production process $e^+e^- \to p + X$ can be
expressed to leading order as
\begin{equation}
\frac{1}{\sigma_{tot}}\frac{d \sigma}{d x_E} =\frac{ \sum\limits_q
\hat{C}_q \left [ D_q (x_E,Q^2)+D_{\bar{q}} (x_E,Q^2) \right ]}
{\sum\limits_q \hat{C}_q}, \label{crosection}
\end{equation}
where  $x_E=2 E_p/\sqrt{s}$,  $s$ is the total center-of-mass
(c.m.) energy squared, $E_p$ is the energy of the produced proton
in the $e^+e^-$ c.m. frame, and $\sigma_{tot}$ is the total cross
section for the process. In Eq.(\ref{crosection}), $\hat{C}_q$
reads
\begin{equation}
\hat{C}_q=e_q^2-2 \chi_1 v_e v_q e_q+ \chi_2 (a_e^2+v_e^2)
(a_q^2+v_q^2),\label{hatC}
\end{equation}
with
\begin{equation}
\chi_1=\frac{1}{16 \sin^2 \theta_W \cos^2 \theta_W}
\frac{s(s-M_Z^2)}{(s-M_Z^2)^2+M_Z^2\Gamma_Z^2},
\end{equation}
\begin{equation}
\chi_2=\frac{1}{256 \sin^4 \theta_W \cos^4 \theta_W}
\frac{s^2}{(s-M_Z^2)^2+M_Z^2\Gamma_Z^2},
\end{equation}
\begin{equation}
a_e=-1,
\end{equation}
\begin{equation}
v_e=-1+4 \sin^2 \theta_W,
\end{equation}
\begin{equation}
a_q=2 T_{3q},
\end{equation}
and
\begin{equation}
v_q=2 T_{3q}-4 e_q \sin^2 \theta_W,
\end{equation}
where $T_{3q}=1/2$ for $u$, while $T_{3q}=-1/2$ for $d$, $s$ quarks.
Moreover $N_c=3$ is the color number, $e_q$ is the charge of
the quark in units of the proton charge,
$\theta$ is the angle
between the outgoing quark and the incoming electron, $\theta_W$
is the Weinberg angle, and $M_Z$ and $\Gamma_Z$ are the mass and
width of $Z^0$.

In Fig.~{\ref{mssy12f1}} we present our results of the proton
production in $e^+e^-$ IA with the three simple options for the
relation between distribution and fragmentation functions
Eq.~(\ref{dfR}). Option 1: $C_v=1$ and $C_s=0$ for $\alpha=0$;
Option 2: $C_v=C_s=1$ for $\alpha=0.5$; and Option 3: $C_v=1$ and
$C_s=3$ for $\alpha=1$. We find that all of these three options
fit rather well the experimental data in a large $z$ range, though
there is some discrepancy at small $z$. From Fig.~1(a) and (b), we
find that there are only small differences in the predictions when
we use the two different choices of quark distributions
parametrizations. Thus our analysis of the relation between
distribution and fragmentation functions is not sensitive to the
available parametrizations of quark distributions which are well
constrained by a vast number of experimental data. All of the
three options support the validity of  the revised Gribov-Lipatov
relation Eq.~(\ref{RGLR}) at $z \to 1$.

\begin{figure}
\begin{center}
\leavevmode {\epsfysize=8cm \epsffile{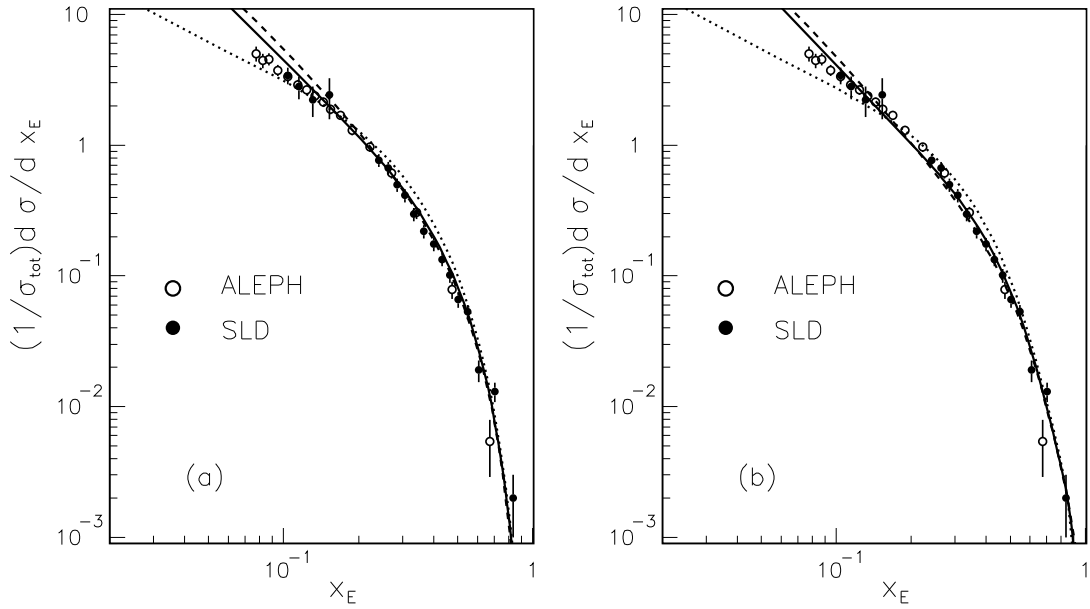}}
\end{center}
\caption[*]{\baselineskip 13pt Predictions for the cross section
of proton production in $e^+e^-$ annihilation. Dotted, solid,
dashed curves correspond to the three options of 1, 2 and 3
respectively, with (a) CTQE and (b) BSB parametrizations of the parton
distributions. The experimental data are taken from
\cite{ALEPH98,SLD99}. } \label{mssy12f1}
\end{figure}

As indicated above, the three options differ in the roles played
by the valence and sea quark contributions in the fragmentation
functions. Option 2 corresponds to a situation with equal
contributions from valence and sea quarks, whereas Option 1 and
Option 3 correspond to the situations with sea suppressed and
enhanced respectively. Though such differences do not show up
significantly in $e^+e^- \to p +X$, they will show up in the
proton/antiproton productions in $p +p \to p (\bar p) + X$, as
will be shown in the following. Thus we can test these three
different options by examining predictions in these processes,
where the roles played by valence and sea quarks are different in
comparison with $e^+e^-$ IA. For the sake of simplicity, we only
adopt the fragmentation functions based on the CTEQ
parametrization of the parton distributions. In fact, the new
facility RHIC at BNL is expected to be running at some high c.m.
energy such as $\sqrt{s}=200~\rm{GeV}$ and
$\sqrt{s}=500~\rm{GeV}$. We thus use these fragmentation functions
to predict the cross section of the proton/antiproton productions
via
\begin{equation}
p + p \to H + X,
\end{equation}
where the produced hadron $H$, is a proton or an antiproton,
respectively. The invariant cross section of the above process can be
expressed as
\begin{equation}
E_H \frac{d^3\sigma}{d^3p_H} = \sum \limits_{abcd} \int_{\bar
x_a}^1 dx_a \int_{\bar x_b}^1 dx_b  f_a(x_a,Q^2)f_b(x_b,Q^2)
D_c^H(z,Q^2) {1 \over \pi z}{ d\hat \sigma \over d\hat t}(ab \to
cd)~. \label{Dsig}
\end{equation}
In Fig.~\ref{mssy12f2}, we present the rapidity distribution of
the cross section for proton production at the c.m. energy
$\sqrt{s}=200~\rm{GeV}$ (left figure (a)) and
$\sqrt{s}=500~\rm{GeV}$ (right figure (b) ), for various values of
the transverse momentum $p_{T}$ of the produced hadron. In
Fig.~\ref{mssy12f3}, the $p_T$ distribution of the cross sections
is shown at the above mentioned two values of the c.m. energy. We
see that the predictions do not show large differences for the
three options. In Fig.~\ref{mssy12f4}-\ref{mssy12f5}, the rapidity
and the transverse momentum distributions of the cross sections
are shown for antiproton production and now we observe significant
differences for the predictions from the three options. This can
be easily understood since antiproton production is sensitive to
the sea quark fragmentation. We thus conclude that antiproton
production in the reaction $p+p \to \bar p + X$ can discriminate
clearly between the three options of quark fragmentation
functions.

\begin{figure}
\begin{center}
\leavevmode {\epsfysize=10cm \epsffile{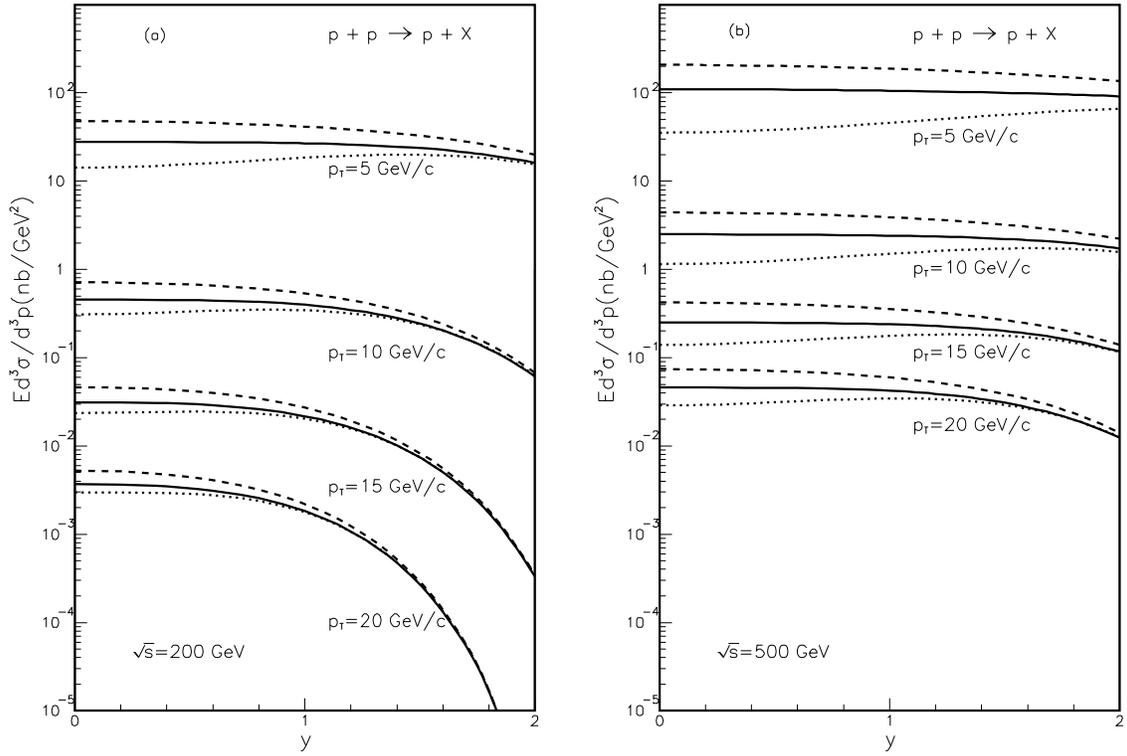}}
\end{center}
\caption[*]{\baselineskip 13pt Predictions of the rapidity
distribution of the cross section of proton production in pp
collision. Dotted, solid, and dashed curves correspond to the
predictions from the three options 1, 2 and 3 respectively, for
c.m. energy  $\sqrt{s}=200$~{GeV}
(a) and $\sqrt{s}=500$~{GeV} (b). }
\label{mssy12f2}
\end{figure}

\begin{figure}
\begin{center}
\leavevmode {\epsfysize=12cm \epsffile{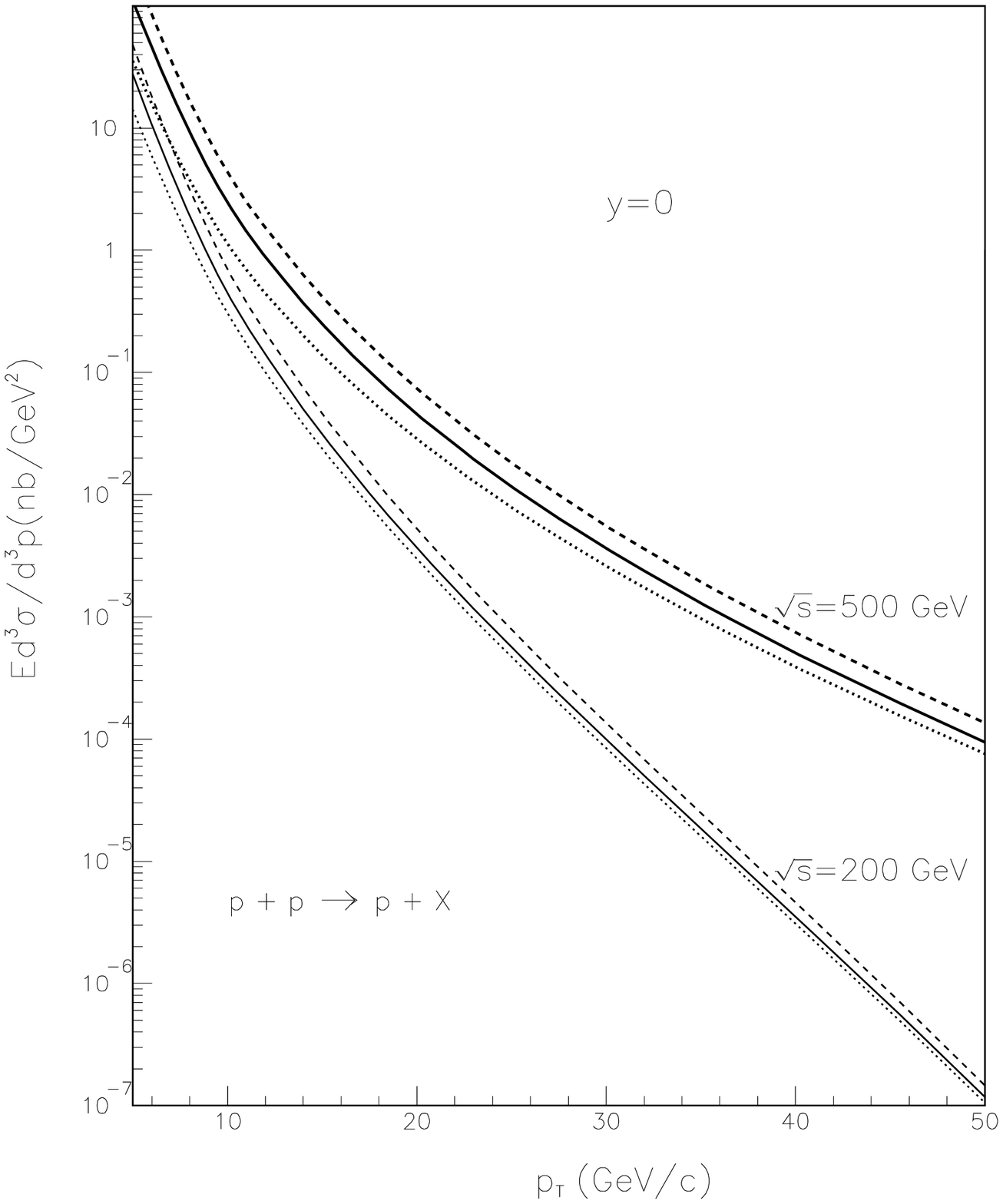}}
\end{center}
\caption[*]{\baselineskip 13pt Predictions of the transverse
momentum distribution of the cross section of proton production in
pp collision. Dotted, solid, and dashed curves correspond to the
predictions from the three options 1, 2 and 3 respectively, for
c.m. energy $\sqrt{s}=200$~{GeV} (a) and
$\sqrt{s}=500$~{GeV} (b).} \label{mssy12f3}
\end{figure}

\begin{figure}
\begin{center}
\leavevmode {\epsfysize=10cm \epsffile{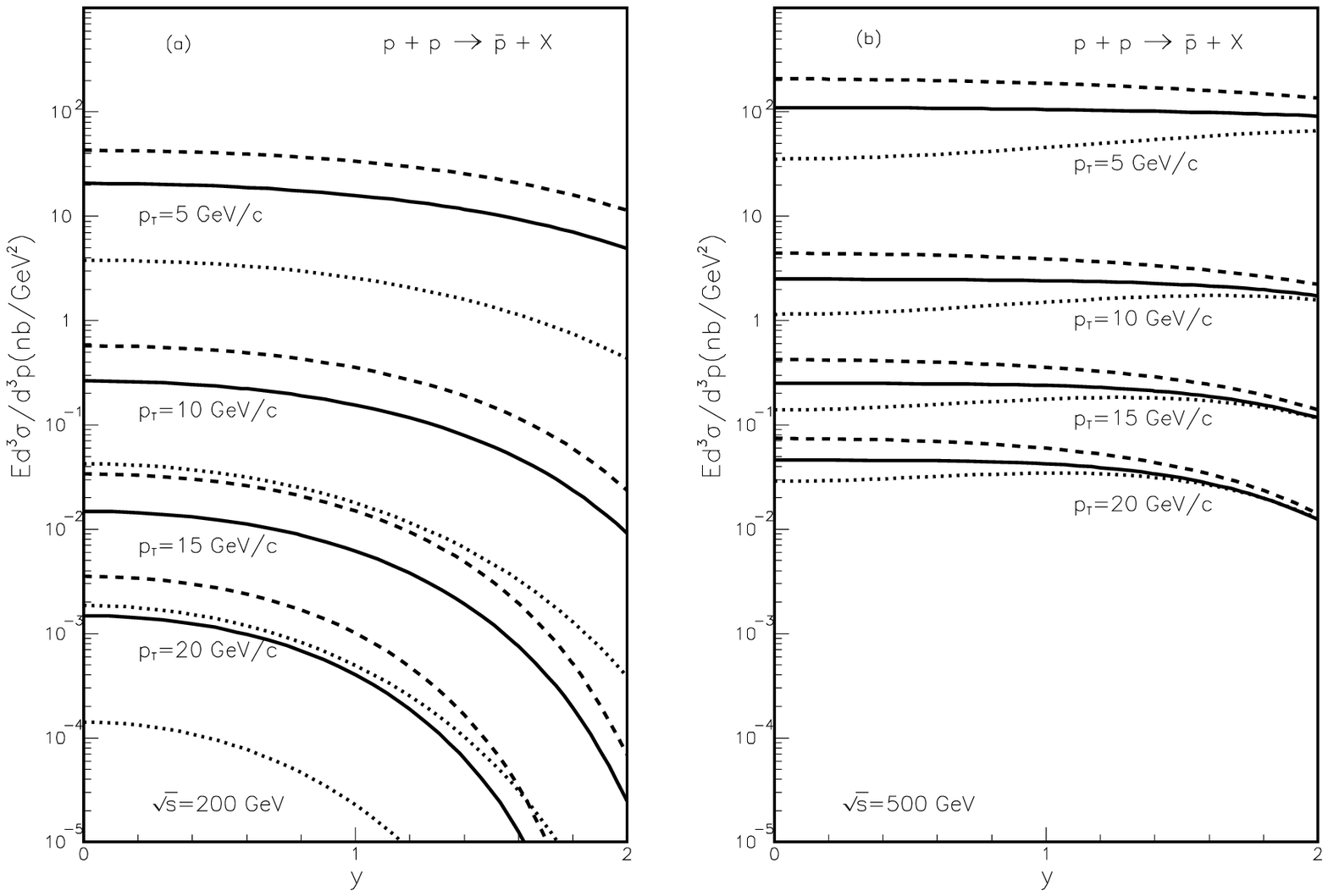}}
\end{center}
\caption[*]{\baselineskip 13pt The same as Fig.~\ref{mssy12f2},
but for antiproton production.} \label{mssy12f4}
\end{figure}

\begin{figure}
\begin{center}
\leavevmode {\epsfysize=12cm \epsffile{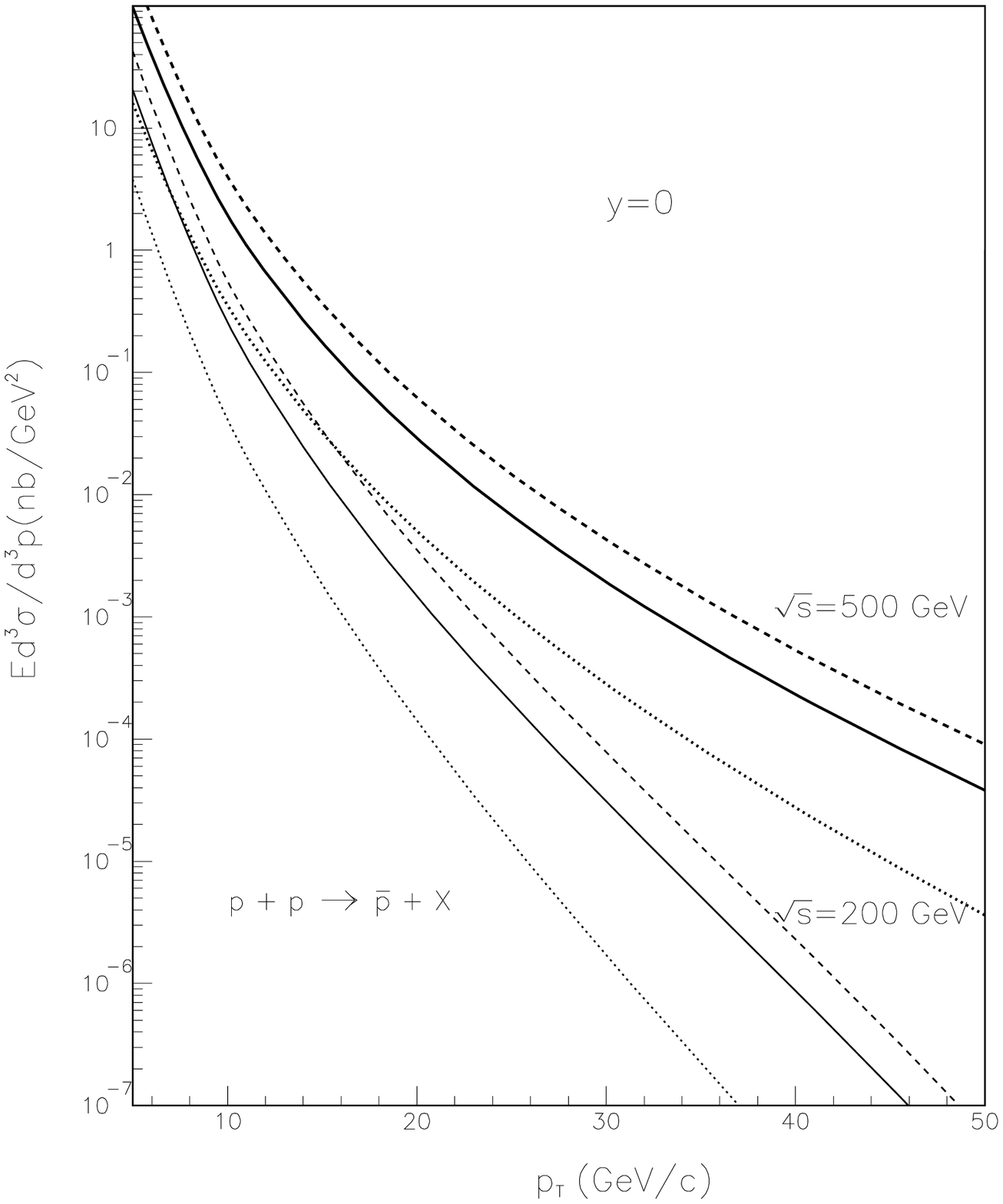}}
\end{center}
\caption[*]{\baselineskip 13pt The same as Fig.~\ref{mssy12f3},
but for antiproton production.} \label{mssy12f5}
\end{figure}

In summary, we have studied the relation between fragmentation
and distribution functions based on known parametrizations of
parton distributions and the available experimental data on proton
production in  $e^+e^-$ IA. With general
relations between fragmentation and distribution functions, we
find three simple options that can provide a good description of the
proton production data in $e^+e^-$ IA. All of the three
options support the revised Gribov-Lipatov relation $D_q(z)=zq(z)$
at $z \to 1$ as an approximate relation, and they differ in the
roles played by valence and sea quarks in the fragmentation. Such
difference can show up especially from antiproton production in
the reaction $p+p \to \bar p + X$, thus they can be tested in future
experiments at RHIC-BNL.

{\bf Acknowledgments: } This work is partially supported by
National Natural Science Foundation of China under Grant Numbers
19975052, 10025523, 90103007 and 10175074, by Fondecyt (Chile)
3990048 and 8000017, by the cooperation programmes
Ecos-Conicyt C99E08 between France and Chile,  and
also by Foundation for University Key Teacher by the
Ministry of Education (China).


\newpage
\begin{thebibliography}{99}


\bibitem{GL}
V.N.~Gribov, L.N.~Lipatov, Phys. Lett.  {\bf B 37} (1971) 78; Sov.
J. Nucl. Phys. {\bf 15} (1972) 675; S.J.~Brodsky, B.-Q.~Ma, Phys.
Lett. {\bf B 392} (1997) 452.


\bibitem{BDM00} V.~Barone, A.~Drago, B.-Q.~Ma,
Phys. Rev. {\bf C 62} (2000) 062201(R).


\bibitem{BRV00} J. Bl\"umlein, V. Radindran, W.L. van Neerven,
Nucl. Phys. {\bf B 589} (2000) 349.


\bibitem{MSSY}
B.-Q. Ma, I. Schmidt, J.-J. Yang,   Phys. Rev. {\bf D 61} (2000)
034017, Phys. Lett. {\bf B 477} (2000) 107; B.-Q. Ma, I. Schmidt,
J. Soffer, J.-J. Yang, Eur. Phys. J. {\bf C 16} (2000) 657; Phys.
Rev. {\bf D 62} (2000) 114009; Phys. Rev. {\bf D 64} (2001)
014007; Phys. Rev. {\bf D 65} (2002) 034004.



\bibitem{DISprocess}
For a review, see, e.g., J.F. Owens, W.-K. Tung, Ann. Rev. Nucl.
Part. Sci. {\bf 42} (1992) 291.


\bibitem{DYprocess}
For a review, see, e.g., P.L. McGaughey, J.M. Moss, J.C. Peng,
Ann. Rev. Nucl. Part. Sci. {\bf 49} (1999) 217.


\bibitem{CTEQ}
CTEQ Collaboration, H.L. Lai {\it et al.}, Eur. Phys. J.  {\bf
C 12} (2000) 375.



\bibitem{BSB01}
C. Bourrely, J. Soffer, F. Buccella, 
Eur. Phys. J.
 {\bf C 23} (2002) 487.


\bibitem{GRV95}
M.~Gl\"uck, E.~Reya, A.~Vogt, Z. Phys.  {\bf C 67} (1995) 433 .


\bibitem{MRST00}
A. D. Martin, R. G. Roberts, W. J. Stiring,
R. S. Thorne, 
Eur. Phys. J.  {\bf C 14} (2000) 133.




\bibitem{ALEPH98}
ALEPH Collaboration, R. Barate {\it et al.},  Phys. Rep. {\bf 294} (1998) 1.


\bibitem{SLD99}
SLD Collaboration, K. Abe {\it et al.},  Phys. Rev.  {\bf D 59}
(1999) 052001.





\end{thebibliography}
\end{document}